# An Experimental and Computational Study of Breaking Wave Impact Forces

T.C. Fu[1], A.M. Fullerton[1], S. Brewton[1], K.A. Brucker[2], and D. Dommermuth[2]
([1] Carderock Division, NSWC, USA;  [2] SAIC, La Jolla, CA, USA)

**ABSTRACT**

The impact forces generated by the impact of a breaking wave are poorly understood. These impulsive hydrodynamic loads to a ship's hull are of short duration relative to ship motions and buoyant wave loads and often result in extremely high pressures. The physics of breaking waves is a poorly understood, complex, multiphase phenomenon involving violent jet sprays, strong free-surface turbulence, air entrainment and bubble generation, all of which interact with the flow field and the adjacent structure.

This paper will describe a set of experiments that were performed, at the Naval Surface Warfare Center, Carderock Division (NSWCCD), in 2006, to measure the hydrodynamic loads of regular non-breaking and focused breaking waves on a 0.305 m x 0.305 m (1.0 ft x 1.0 ft) square plate and discuss the results of this study. The paper will also discuss Computational Fluid Dynamics (CFD) code predictions of breaking waves and wave impact loads. The CFD code utilized in this study is Numerical Flow Analysis (NFA).

**INTRODUCTION**

Marine platforms, whether they be moving (ships, submarines, etc) or stationary (seawalls, breakwaters, platforms, etc) are all exposed to loads from breaking wave impacts (see Figure 1), and there many documented cases of wave impact damage (Buchner and Voogt, (2004) for example).

Though much work has been done in this area, the answer to the simple question of "what is the maximum wave impact pressure possible in the field?" is still unclear. Chan and Melville (1984, 1987, 1988, & 1989) investigated the force of plunging breakers on flat plates and vertical cylinders, showing breaking wave impact pressures as high as $10\rho c^2$, where $\rho$ is the density of water and $c$ is the wave celerity. Experimental results from Zhou, Chan and Melville (1991) found similar results. The pressures measured in Chan's and Melville's experiment are "localized to within the zone of impact" (1989). Field data collected by Bullock and Obhrai (2001) shows pressures of over 383 kilopascals (8000 lbs/sq ft.) on a breakwater for an incident wave height of 3 m (10 feet), while Peregrine (2003) has measured peak pressures an order of magnitude higher than this value in the laboratory. It has been hypothesized that these extremely high pressures may be due to the refraction and focusing of pressure waves caused by large spatial variations in the velocity of sound associated with different levels of voids ratio. Much more work is needed, focusing on the physics of breaking waves and the mechanisms which produce these extreme impulse pressures.

The low velocity of sound in the air-water mixture present in a breaking wave may, in fact, be a major factor in determining the maximum pressure, but this is still unclear. The acoustic limit, or water hammer pressure, was thought to be too high to be realizable in the field, but recent works by Peregrine & Thais (1996) and Peregrine (2003) indicate that this may not be true when aeration is taken into account. Indeed, even higher pressures may be possible due to the refraction and focusing of pressure waves caused by large spatial variations in the velocity of sound associated with different levels of void ratio. Field and laboratory measurements have recorded instances of sub-atmospheric pressure during coastal wave impact (Oumeraci et al. 1993; Hattori et al. 1994; Bullock et al. 2005; Bullock et al. 2007), indicating the potential presence of fluid cavitation, which may cause additional damage to the structure. The presence of  entrained air and cavitation presents additional complexity to the scaling of wave impact loads.

The computational study of waves, particularly breaking waves, and fluid-structure interaction are fairly new fields. Efforts tend to fall into four camps.

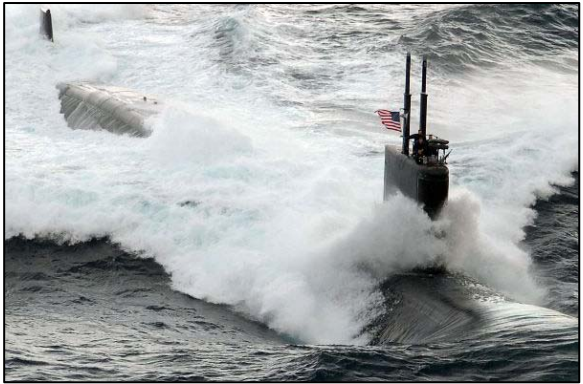
a) Surfaced Submarine

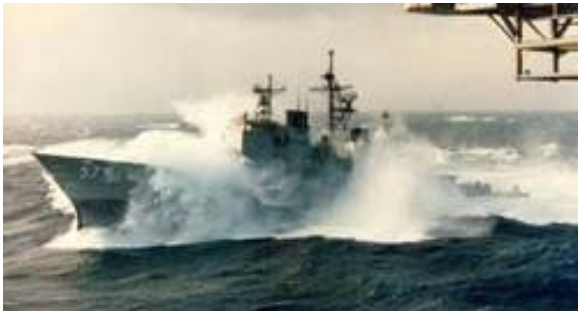
b) Surface Ship

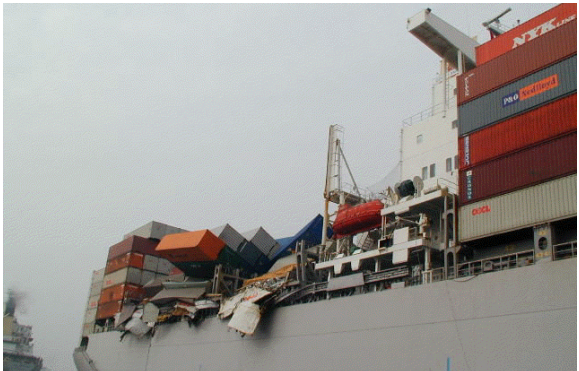
c) Wave impact damage to a container ship

**Figure 1:** Examples of large wave impact loading, a) surfaced submarine, b) surface ship in a seaway, and c) damage to the deck containers on a containership.

One area of study is focused on using non-linear potential flow codes and boundary element methods to study the impact of waves on off-shore structures and break waters (Corte & Grilli, 2006 and Hur et al, 2004). While this method shows promising results for computing the impact forces of non-breaking waves, the very nature of the potential method may limit the codes' ability to accurately replicate a breaking wave. The second area relies on RANS codes to study hydrodynamic forces on breakwaters and other shallow water structures. This approach takes advantage of the bottom slopes to create breaking waves as they approach a fixed structure. Several methods have been used to simulate these events including: a level-set model to look at wave-breaking processes in shallow water (Gomez et al, 2002) and an application of Cointe and Tullin's (1994) theory of steady breakers to look at spilling waves on a submerged hydrofoil (Rhee & Stern, 2002). Still a third research area, again using RANS, has focused on ship generated waves. Waves created by the ship have been modeled computationally, and the tools have been able to capture breaking bow wakes and related secondary free surface flows. Gorski et al (2006), Miller et al (2006) have demonstrated this capability for a number of High Speed Sealift surface ships. The fourth area is using a smoothed particle hydrodynamic (SPH) method – a mesh-free particle approach to track large deformations in the free surface. This method has shown some success in analyzing spilling and plunging waves on slopes with limited validation.

Within the body of computational work, few efforts appear to be focused on understanding the physics of wave impact on structures. Clauss, (Technical University Berlin) has a significant body of work in both computational and experimental studies with breaking waves. This includes Clauss et al (2005), which coupled several commercial RANS solvers with an in-house potential finite element code called WAVETUB to determine the wave impact forces in varying seas. The focus of this work was not only on creating an efficient method of simulating breaking waves, but also on providing a method to study the instantaneous velocities and accelerations at the wave crest. Additionally, Lugni et al (2006) built upon Faltinsen's early work on wave slamming (Faltinsen (2000) and Faltinsen et al (2004) for example) and looked at the role of flip through.

Because computational codes are just starting to address the issues related to breaking waves, their accuracy in predicting impact forces has yet to be fully explored. Several efforts are underway to create accurate and consistent breaking wave trains and predict their impact forces, which can be derived from the transient pressure distribution. To support ongoing CFD efforts and to provide engineering guidance to the US Navy, the Hydromechanics Department at NSWCCD funded a canonical wave impact test to begin to illuminate the number of issues related to breaking wave impact forces and their prediction, while the Office of Naval Research sponsored the CFD effort.

# EXPERIMENT

In 2006, a set of experiments was conducted at NSWCCD to measure the hydrodynamic loads of focused breaking and regular non-breaking waves on a 0.305 m x 0.305 m x 0.051 m (1 ft x 1 ft x 0.17 ft) aluminum flat plate. This work complemented a similar effort which measured non-breaking wave impact loads (Fullerton & Fu, 2007). The plate was suspended from Carriage 5 and was held stationary in the High Speed Tow Basin approximately 45.7 m from the wavemaker (Figure 2a). The setup was chosen to simulate a 2-dimensional wave impact problem, and Figure 2b shows a schematic diagram of the flat plate that was tested. Three-component force measurements were collected using a Kistler gage mounted to the plate. Incoming waves were measured using six Senix TS-15 distance sensors sampling at 10 Hz. These instruments are non-contact, ultrasonic instruments for measuring distances in air, with a capability to measure distances from 5 cm to 11 m (2 inches to 37 ft), with accuracies of 0.1% at 20 Hz. A visual record of the wave impacts was collected using both standard and high-speed video cameras and digital recorders (see Figure 2a).

The locations of the six Senix sensors are as follows: three sensors were located along the centerline of the plate at 0.076 m (3 inches) in front of the plate, 0.47 m (18.5) inches in front of the plate, and 3.66 m (12 feet) in front of the plate. Two additional Senix sensors were located off the centerline of the plate, with one at 0.08 m (3 inches) forward and 0.46 m (18 inches) off the centerline and one at 0.46 m (18 inches) forward and 0.46 m (18 inches) off the centerline. The final sensor was placed in line with the plate longitudinally along the

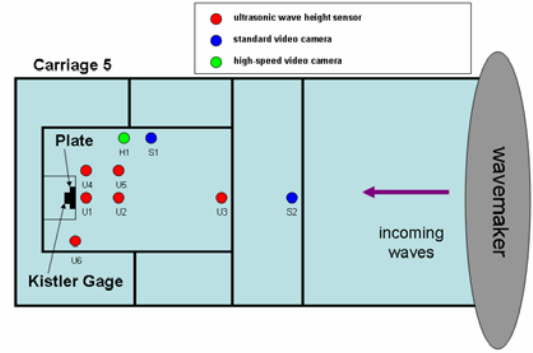

a) Experimental set-up

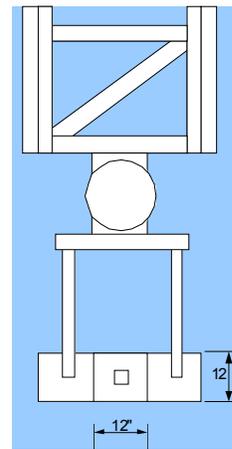

b) Instrumented plate and supporting structure

**Figure 2:** a) Sketch of the experimental set-up showing the carriage, plate, wave height sensors and wavemaker and b) the plate showing the instrumented center section.

tank, but off centerline to measure the breaking wave in absence of the plate (see Figure 2a).

Impact data was acquired at three different levels of plate submergence (full, half and none), and three different plate angles (0°, +45° toward the incoming wave, -45° away from the incoming wave), which are shown in Figure 3. Data were collected for about 30 seconds from the time the plate encountered the first wave. Table 1 summarizes the conditions tested during the experiment. Each condition was tested two to three times to ensure repeatability. The water depth during testing was about 4 m (13 feet).

**Table 1: Conditions tested.**

| Wave Amplitude (cm) | Breaking Or Regular | Plate Tilt Angle (degrees) | Submergence Level |
|---|---|---|---|
| 25.4 | Breaking | 0, +45, -45 | none, half, full |
| 19.1 | Breaking | 0, +45, -45 | none, half, full |
| 7.6 | Regular | 0, +45, -45 | none, half, full |
| 10.2 | Regular | 0, +45, -45 | none, half, full |
| 15.2 | Regular | 0, +45, -45 | none, half, full |

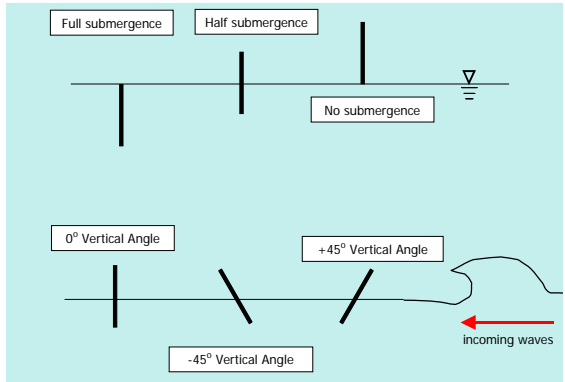

**Figure 3:** Plate submergence levels and angle orientations during testing

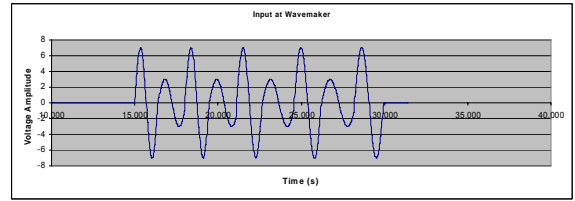

**Figure 4:** Wavemaker voltage input for breaking wave.

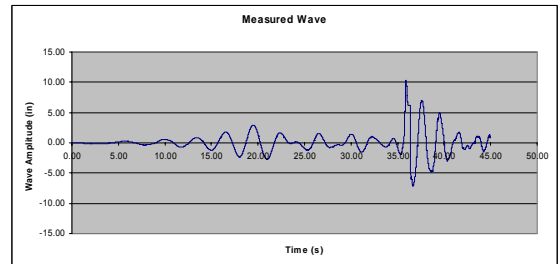

**Figure 5:** Wave measurement near the plate

## EXPERIMENTAL RESULTS

### Impacting Waves

A breaking wave was generated using a combination of non-breaking waves of various frequencies. The breaking waves were generated by sending an external voltage signal made up of 9 waves of varying frequencies (see Figure 4). The shortest waves were sent out first, with increasingly longer waves being sent out in sequence. Since the speed of an individual wave is proportional to the square root of its wavelength, a shorter wave will travel more slowly than a longer wave, and all the waves will meet at some distance from the wavemaker. These individual waves were chosen to combine approximately 61 m (200 feet) from the wavemaker and form a breaking wave. The waves that were used to make the breaking wave are the larger waves in Figure 4 (greater than 6 volts); the smaller waves were inserted to create a smooth input signal so the wavemaker did not have to come to an abrupt halt between large waves. The 0.25 m (10 inch) breaking wave was created using the voltage input shown in Figure 4 and a blower RPM of 1600. Figure 5 shows the wave measurement near the plate and the typical shape of the resultant wave. The smaller breaking wave that was tested was generated using the same voltage input and a blower RPM of 1100. Regular waves were generated through the specification of blower speed (RPM) and frequency (Hz).

### Impact Forces

Figure 6 shows the normal force (Fx) and vertical force (Fz) on the plate with no tilt angle (0 degrees) for non-breaking waves. Fx is positive in the direction of the incident wave, and Fz is positive up. Panel 1 (top) shows the force on the plate with no immersion, panel 2 shows the force on the plate half immersed, and panel 3 shows the force on the plate when fully immersed. The force gage was zeroed out before each run, so the forces shown are due only to the wave impact on the plate. Panel 1 shows the flat plate with no immersion only feels a positive impact force in the direction of wave propagation. When the plate is half submerged (panel 2), it begins to experience some negative force and for the fully submerged plate, the maximum positive and negative values are almost the same.

Figure 7 shows still images of a wave with amplitude = 0.305 m and wavelength = 6.1 m, impacting a tilted (45 degrees forward) flat plate with no submersion. Figure 7a shows the plate before impact of a non-breaking wave, and Figure 7b shows the plate during the wave impact, where the wave breaks and overtops the plate. Figure 8 shows a sample of the measured data for a breaking wave: 8a) wave amplitude and 8b) the wave impact force.

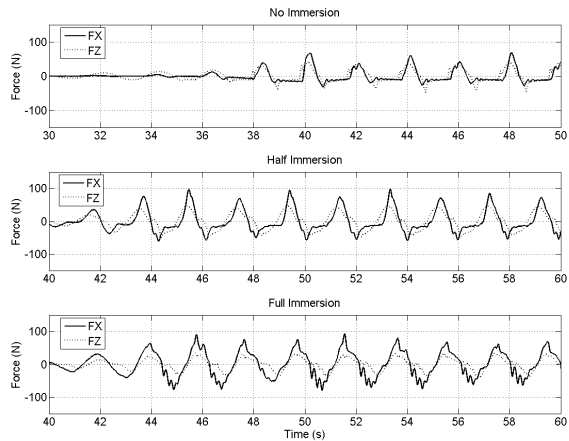

**Figure 6:** Forces from 0.305 m (12 inch) amplitude, 6.1 m (20 foot) wavelength non-breaking wave on a flat plate with 0 degree vertical aspect angle. Panel 1 (top) shows the force on a plate with no immersion, panel 2 shows the force on a plate with half immersion, and panel 3 shows the force on a plate with full immersion.

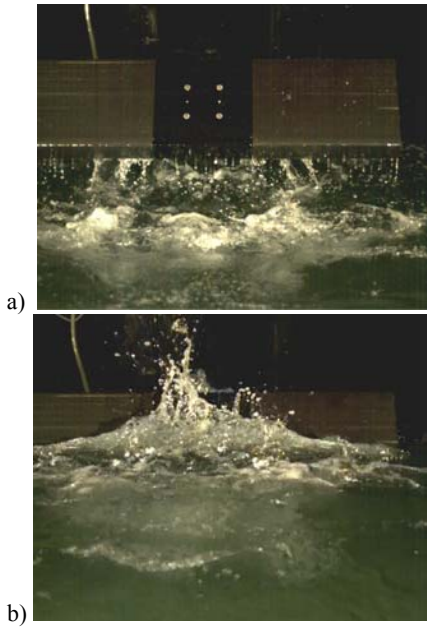

**Figure 7:** Still images of a wave with amplitude = 0.305 m and wavelength = 6.1 m, impacting a tilted (45 degrees forward) flat plate with no submersion: a) before wave impact and b) during impact.

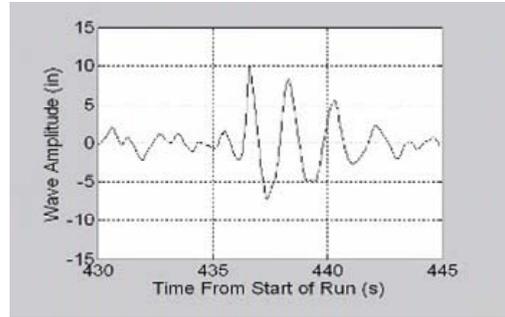

a) Wave Amplitude (in)

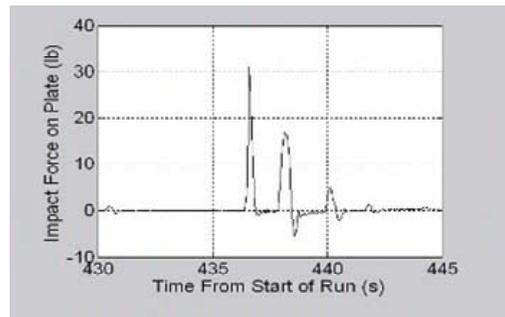

b) Impact Force (lbs)

**Figure 8**: Sample wave amplitude and impact force data for a typical wave impact.

**Discussion**

Figure 9 shows the breaking wave amplitude and the impact force (Fz) normal to the plate with 0° angle for six separate impact events. From this figure, it can be seen that the impact force varies significantly, even while wave height remains similar for each impact. Figure 10 shows the average impact forces for the 0.25 m amplitude breaking wave for the range of plate conditions tested. Individual impacts can vary by as much as 133.5 newtons (30 lbs). Note that the no submergence cases yield the highest forces, as expected. Plate angles of 0° and 45° toward the incoming waves generated higher forces than the condition with the plate angled away from waves. Figure 11 shows a comparison of all the forces normal to the plate from breaking and non-breaking waves for vertical plate angle of 0°. This plot shows that some breaking waves follow the trend of the non-breaking waves, however some breaking waves can be more than twice as high as the trend predicted by the non-breaking waves. This plot also shows that there is much more variation in the forces measured from breaking waves. Wave breaking is a random process, and is dependent on the amount of air entrained by the breaking wave, so even if the breaking height is similar to another case, the impact forces can vary significantly.

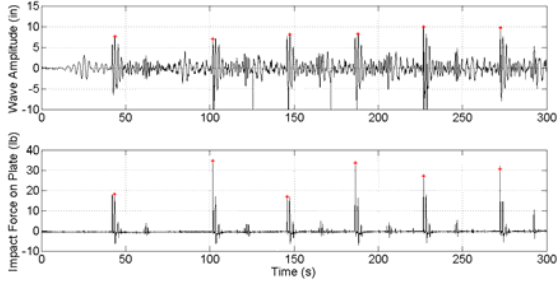

**Figure 9:** Time series of the wave amplitude and impact force for 0.25 m (10 inch) breaking wave on 0° vertical plate angle with no submergence

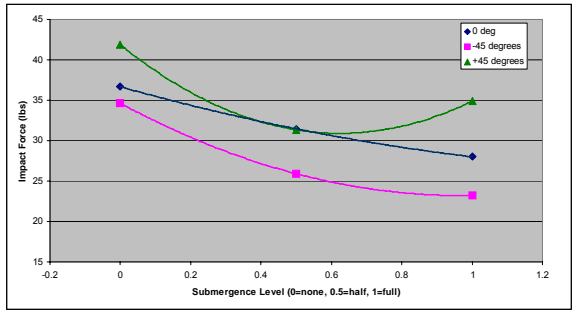

**Figure 10:** Average impact forces for the range of conditions tested

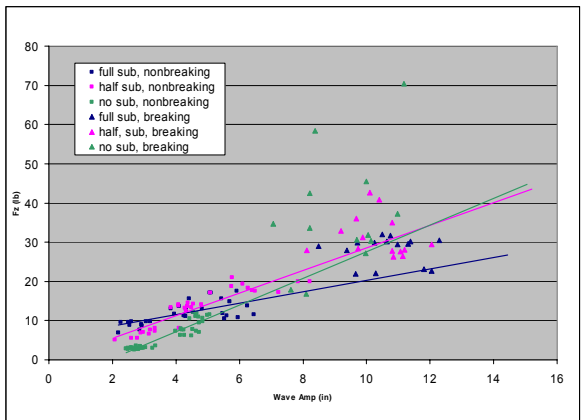

**Figure 11:** Comparison of all breaking/non-breaking waves for vertical plate angle of 0°

Much greater variation in impact forces was observed from the breaking wave cases than from the non-breaking waves, with the individual breaking wave impact forces varying by as much as 70% of the average impact value. The plate with a 0 degree vertical angle and 45 degree vertical angle toward the incoming waves generated higher forces than the orientation with the plate angled away from the waves.

**NUMERICAL COMPUTATIONS**

**Numerical Formulations**
O'Shea, Brucker, Dommermuth, & Wyatt (2008) provide details of the NFA formulation. Here, we highlight the formulation of a wavemaker. Dommermuth, Yue, Lin, Rapp, Chan & Melville (1988) compare the predictions of boundary-integral equation method (BIEM) to laboratory measurements of a plunging breaking wave. The numerical predictions compare well with laboratory measurements up to the point of wave impact. Unlike BIEM, a volume-of-fluid formulation like NFA is capable of predicting fluid motion beyond wave impact.

The motion of the wavemaker is prescribed in terms of a Fourier series:

$$U(t) = \sum_{i=1}^{N} U_n \cos(\omega_n t - \theta_n) \qquad (1)$$

where $U(t)$ is the measured time history of the horizontal velocity of the wavemaker, $U_n$ is the Fourier amplitude, and $\omega_n$ and $\theta_n$ are respectively the Fourier frequencies and phases for each harmonic. The Fourier amplitudes, frequencies, and phases for N = 72 modes are provided in Dommermuth et al. (1988) for a plunging-breaker experiment. As in Dommermuth et al. (1988), the length and velocity scales are respectively normalized by d and $\sqrt{gd}$, where d is the water depth and g is the acceleration of gravity.

A no-flux boundary condition is imposed on the surface of the wavemaker using a finite-volume technique. A signed distance function $\psi$ is used to represent a piston-type wavemaker. $\psi$ is positive in the fluid ahead of the wavemaker and negative behind the wavemaker. The magnitude of $\psi$ is the minimal distance between the position of $\psi$ and the surface of the wavemaker. $\psi$ is zero on the surface of the wavemaker. Details associated with the calculation of $\psi$ for general geometries are provided in Sussman & Dommermuth (2001) and Dommermuth, O'Shea, Wyatt, Sussman, Weymouth, Yue, Adams & Hand (2006).

In our two-phase formulation, the piston-type wave-maker spans the entire domain from the bottom of the water to the top of the air. Since both the air and water are incompressible in our volume-of-fluid formulation, the net flux integrated over the surfaces of the computational domain should be zero. This constraint is imposed in the present formulation by prescribing sources and sinks in the air ahead of the wavemaker that cancel out the flux of fluid that is induced by the motion of the wavemaker.

**Numerical Results**
We consider the impact of a plunging breaker with a flat plate in two parts: pre-impact and impact. Two-dimensional numerical simulations are used to simulate the evolution of the wave packet as it moves down the tank prior to impact, and three-dimensional simulations are required at impact. The two-dimensional simulations are used to initialize the three-dimensional simulations.

A pneumatic wavemaker is used in the NSWCCD model basin to generate a packet of waves that coalesce through the effects of dispersion. The free-surface elevation is measured at several points in the basin. This wave-probe data can be used to initialize a numerical simulation based on linear wavemaker theory. Dean and Dalrymple (1991) provides a derivation of linear wave theory that can be used to calculate the transfer function between free-surface elevation and wavemaker motion. In the final step, the calculated wavemaker motion can be used as input to a numerical wavemaker. The actual wavemaker used in the basin and the numerical wave-maker do not have to match. The point is to numerically generate a wave packet that replicates the experiments. This makes it possible to use a much shorter tank in the numerical simulations than had been used in the actual laboratory experiments. In addition, using this procedure, the motion of the actual laboratory wavemaker is not required.

Although we did not complete our numerical studies of the NSWCCD experiments all the way through to wave impact, we have performed a pre-impact study of a packet of waves that forms a plunging breaker and a study of the forces acting on a sphere impacting the free surface. These studies are described in the next two sections.

**Analysis of a Plunging Breaking Wave**
Two simulations with different grid resolutions of a wavemaker experiment have been performed. The data for the laboratory experiment is provided in Dommermuth et al. (1988). The numerical formulation is outlined earlier in this paper. Up to the point of wave impact, the experiments are two-dimensional. NFA is a three-dimensional formulation, so for these preliminary numerical simulations, the tank is modeled as a narrow three-dimensional slice. A coarse-resolution simulation with 1344x2x128= 344,064 grid points and a medium-resolution simulation with 1344x2x192=516,096 grid points have been performed. The coarse and medium simulations respectively use 42 and 63 subdomains. For each simulation, the subdomain is assigned to a single node on a Cray XT3. The grid spacing is constant for the coarse simulation ($\Delta$=0.015625). For the medium-resolution simulation, grid stretching is used near the wavemaker at x =0, the mean free-surface, and the point of wave impact. The minimum grid spacing for the medium-resolution simulation is about half the coarse simulation ($\Delta$=0.00749). For the coarse-resolution simulation, the length, water depth, and air height of the computational domain are respectively 20.5, 1.0, and 1.0. For the medium-resolution simulation, the length, water depth, and air height of the computational domain are respectively 20.5, 1.0, and 2.0. For these simulations, the non-dimensional time steps are $\Delta t$ =0.004 & 0.002 and the numerical simulations run 13,000 and 26,000 time steps for the coarse and medium resolution simulations, respectively.

Figures 12 (a-f) compare medium-resolution predictions of the free-surface elevation to experimental measurements for six different positions downstream of the wavemaker. The agreement is excellent. The numerical simulations stopped just before wave impact because at that point, the two-dimensional simulations should be used to initialize three-dimensional simulations.

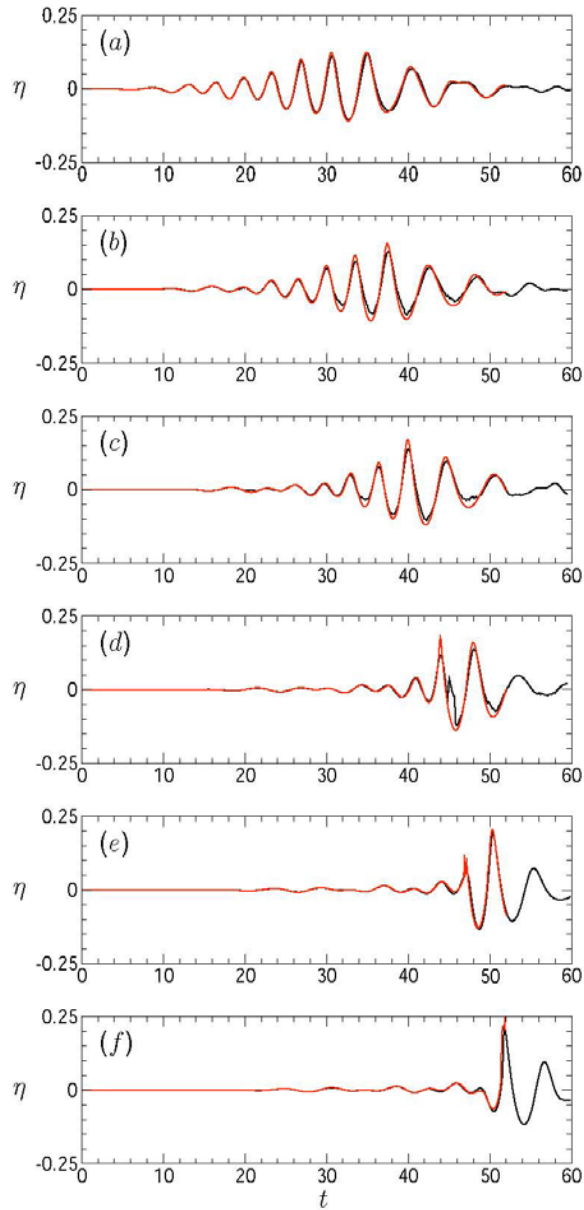

**Figure 12:** Free-surface elevations comparing numerical prediction (red line) to experimental wave-probe measurements (black line) as a function of time at distances from the wave maker of (a) x=3.17, (b) x=2.00, (c) x=6.67, (d) x=9.17, (e) x=10.83 and (f) x=11.83. Results are for the medium resolution case.

Figures 13 (a-g) compare medium-resolution predictions of water-particle velocities to experimental measurements for various locations above and below the mean free surface. The experimental data is measured using a laser anemometer. When the measuring point is above the free surface, the laser anemometer measures zero. The numerical predictions are based on a two-phase formulation. As a result, the numerical predictions include the flow in the water and the air. In the NFA formulation, the component of velocity that is normal to the free surface is continuous, whereas the tangential component of velocity is discontinuous. The agreement is excellent. Figure 13 (a & b) shows a drop out where a trough passed beneath the laser anemometer. At these locations, the NFA predictions are discontinuous for the u component and continuous for the w component. In Figure 13 (c & d) the laser anemometer is aimed above the mean free surface. The laser anemometer only measures something when wave crest crosses above the measuring point. The subsequent plots show similar behavior that depends on the location of the measuring point relative to the free surface. As before, the numerical simulations stopped just before wave impact because at that point, the two-dimensional simulations should be used to initialize three-dimensional simulations.

Figures 14 (a-f) compare the coarse and medium-resolution simulations during the time of the plunging event. The coarse simulation breaks slightly sooner than the medium simulation. The BIEM simulations that are reported in Dommermuth et al. (2008), which are coarser than either of the present simulations, break even sooner near time $t = 51.55$. Along with the numerical predictions, experimental data from the capacitance wave-probe located at $x = 11.83$ is also plotted. Initially, for $t = 51.6$ and $t = 51.68$, it is not clear how the capacitance wave-probe would respond to a multi-valued free surface. After $t = 51.68$, the agreement between predictions and measurements is very good. These simulations of a plunging wave are the first step toward making quantitative comparisons to wave impact studies in the laboratory from the time of wave generation to the time of wave impact. In the next section, we perform numerical simulations of a sphere impacting with a free surface as an intermediate step toward the full wave-impact problem.

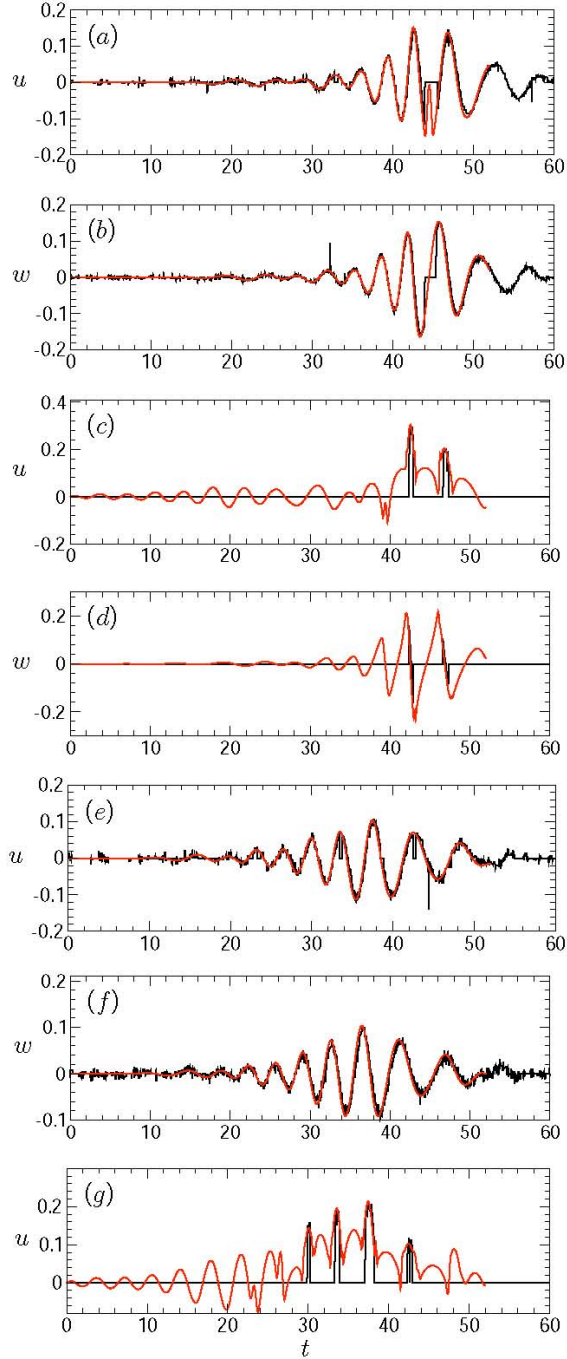

**Figure 13:** Water-particle velocities according to numerical prediction (red line) compared to measurements (black line) using a laser-Doppler anemometer as a function of time. (a) u (8.33, 0.10,$t$), (b) w (8.33, 0.10,$t$), (c) u (8.33, 0.10,$t$), (d) w (8.33, 0.10,$t$), (e) u (5.00, $^-$0.25,$t$), (f) w (5.00, 0.25,$t$), (g) u (5.00, 0.067,$t$) Results are for the medium resolution case.

**Sphere Impact**

Two simulations with different grid resolutions of a sphere impacting a free surface have been performed and compared to laboratory measurements. The data for the laboratory experiment is provided in Laverty (2003). The sphere drops vertically toward the free surface. Length and velocity scales are respectively normalized by the diameter and the velocity of the sphere. The Froude number is 6.325, based on diameter and velocity. Symmetry is imposed across the x=0 and y=0 planes. A coarse-resolution simulation with 256x256x256= 16,777,216 grid points and a medium-resolution simulation with 512x512x512=134,217,728 grid points have been performed. The coarse and medium simulations respectively use 64 and 256 subdomains. For the each simulation, each subdomain is assigned to a single node on a Cray XT3. Grid stretching is used along the cartesian axes to cluster points near the sphere impact region. The smallest grid spacing for the coarse simulation is 0.00310 near the sphere and the largest is 0.0257 near the edges of the domain. The medium resolution simulation is twice as fine. The length, width, water depth, and air height of the computational domains are respectively 2, 2, 2, and 0.6. For these simulations, the non-dimensional time steps are Δt =0.0004 & 0.0002 and the numerical simulations run 750 and 1500 time steps for the coarse and medium resolution simulations, respectively. The spheres are started from rest from a point slightly above the free surface. The centers of the spheres are initially located at z =0.55. A adjustment procedure is used to bring the spheres up to full speed. The adjustment period is 0.5 (Dommermuth et al. 2008). The coarse and medium-resolution simulations respectively took 3.2 and 13.5 wall-clock hours of cpu time.

Figure 15 compares the results of the coarse and medium simulations to laboratory measurements. The vertical force is plotted versus the depth of immersion normalized by the radius of the sphere. The agreement is fair. The numerical predictions rise at the same rate as the experiments, but the peak forces as predicted by numerics are less than measurements. One possible contributing factor is that the experimental forces are deduced from measurements of the sphere's velocity in combination with approximations based on momentum theory. We conjecture that this blend of measurements and theory break down as the depth of immersion increases. Additional studies are required to quantify the present numerical scheme's ability to predict impact loading. Also, we need to assess whether the effects of compressibility require modeling.

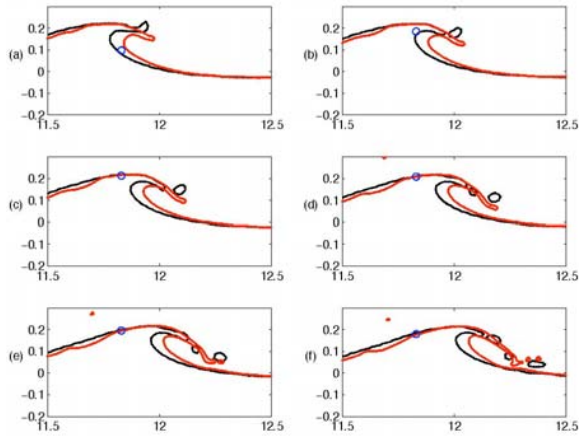

**Figure 14:** Details of breaking event at (a) t=51.6, (b) t=51.68, (c) t=51.76, (d) t=51.84, (e) t=51.92, and (f) t=52. Results compare coarse (black line) and medium-resolution (red line) simulations. The blue circular symbols denote probe measurements at x = 11.83.

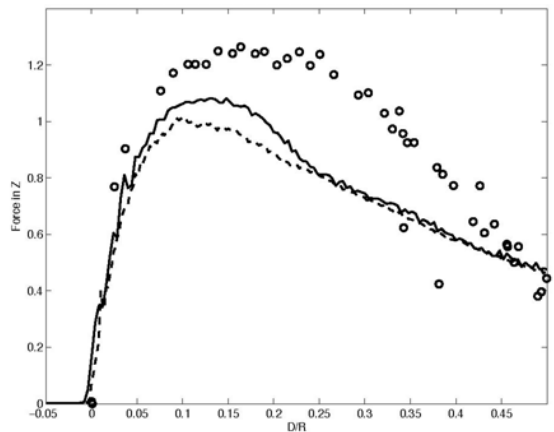

**Figure 15:** Impact loads acting on a sphere as function of the depth of immersion. Experimental measurements (circular symbols), coarse simulations (dashed line), and fine simulations (solid line) are plotted. t =0 is time of impact.

## CONCLUSIONS

The forces generated from breaking and non-breaking wave impact with a flat plate is the focus of an ongoing experimental and computational effort. The testing to date has demonstrated the ability to generate a breaking wave in the high-speed basin at NSWCCD and measure the resultant forces. The data showed :

a) More variation in impact forces measured from breaking waves compared to the non-breaking waves.
b) For breaking waves, the no submergence case yielded the largest forces.
c) The opposite is true for the non-breaking waves, the fully submerged case yielded the largest forces.
d) Some breaking waves followed the trend of the non-breaking waves, but some breaking wave impacts created forces twice as high.
e) Impact forces can vary by as much as 133.5 newtons (30 lbs) or up to 70% of the average impact value.
f) The cases where the plate was vertical ($0^o$ aspect angle) and $45^o$ toward the incoming waves generated higher forces than the case with the plate angled away from the waves.

Additionally, the CFD effort successfully demonstrated the current capability to accurately simulate large breaking waves and predict breaking wave impact loads.

Continuing efforts are focused upon predicting and measuring the wave impact forces of more complex geometries. Figures 16 and 17 show a sample of a more realistic wave impact problem. The Advanced Swimmer Delivery System (ASDS) is shown mounted on a surfaced submarine (Figure 16). Figure 17 shows a sample CFD simulation of this problem.

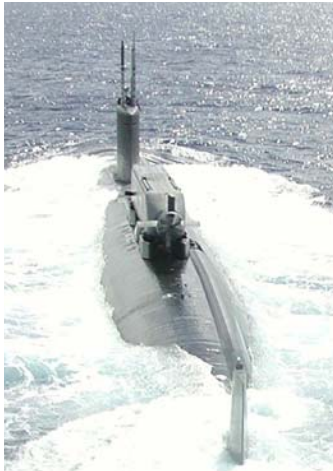

**Figure 16:** Typical wave impact problem of interest. The Advanced Swimmer Delivery System (ASDS) mounted to a surfaced submarine.

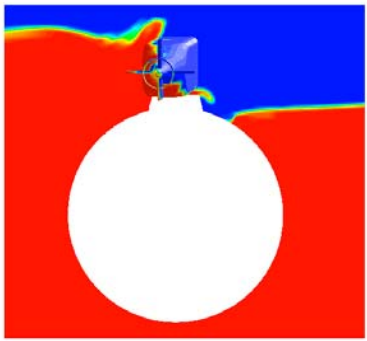

**Figure 17:** Sample CFD simulation of a wave impacting the ASDS mounted on a surfaced submarine.

## ACKNOWLEDGEMENTS

This work was sponsored by the Office of Naval Research, NAVSEA and NSWCCD. The SAIC work was performed under ONR contract number N00014-07-C-0184. Dr. Patrick Purtell is the ONR program manager. This work is also supported in part by a grant of computer time from the DOD High Performance Computing Modernization Program (http://www.hpcmo.hpc.mil/). The numerical simulations have been performed on the Cray XT3 at the U.S. Army Engineering Research and Development Center (ERDC).

The experimental work could not have been accomplished without the efforts of James Rice (NSWCCD, Code 5600), Donnie Walker (NSWCCD, Code 5800) and Mary Lee Pence (CSC) and the encouragement and support of Dr. Edward Ammeen (NSWCCD, Code 5600), Matthew King (SEA 05Z11) and Jack Lee (SEA 05Z11).